\documentstyle[aps,prb,floats,epsf]{revtex}

\begin{document} 

\draft

\wideabs{

\title{Interlayer exchange coupling: Preasymptotic corrections}

\author{P. Bruno\cite{e-mail}}

\address{Max-Planck-Institut f\"ur Mikrostrukturphysik, Weinberg 2, D-06120 
Halle, Germany}

\date{7 January 1999}

\maketitle

\begin{abstract}
In the asymptotic limit, the interlayer exchange coupling decays as $D^{-2}$,
where $D$ is the spacer thickness. A systematic procedure for calculating
the preasymptotic corrections, i.e., the terms of order $D^{-n}$ 
with $n \ge 3$,
is presented. The temperature dependence of the preasymptotic corrections is 
calculated. The results are used to discuss the preasymptotic corrections for 
the Co/Cu/Co(001) system.
\end{abstract}

\pacs{Published in: Eur. Phys. J. B {\bf 11}, 83--89 (1999)}

} 

\section{Introduction}

Since its discovery in 1990,\cite{ Parkin1990} the phenomenon of oscillatory 
interlayer exchange coupling has been the subject of an intense experimental 
and theoretical activity.\cite{ Heinrich1994, Slonczewski1994} Its mechanism 
is now well understood and can be attributed to the quantum size effect due 
to spin-dependent confinement of electrons in the non-magnetic spacer 
layer.\cite{ Edwards1991, Bruno1993, Stiles1993, Bruno1995}

A remarkable result of the theory of interlayer exchange coupling is that it 
becomes particularly simple in the limit of large spacer thickness $D$
(asymptotic regime): (i) the periods of oscillations versus spacer layer 
thickness are uniquely determined by stationary spanning vectors of the bulk 
Fermi surface of the spacer material,\cite{ Bruno1991} (ii) the oscillation
amplitude has a universal $D^{-2}$ decay\cite{ Yafet1987} (except for special
nested Fermi surfaces which give rise to a $D^{-3/2}$ or
$D^{-1}$ decay law\cite{ Bruno1992}), (iii) the amplitudes and phases of the 
oscillatory components are determined respectively by the modules and arguments 
of the spin asymmetry of the (complex) electron reflection coefficients on the 
ferromagnetic layers bounding the spacer layer.\cite{ Bruno1993, Stiles1993, 
Bruno1995} 

Although these simple rules stricly speaking hold only in the limit 
of infinite 
spacer thickness, they proved to be extremely successful in explaining and 
predicting the interlayer exchange coupling observed experimentally for spacer 
thicknesses down to less than 10 atomic layers (AL).\cite{ Unguris1994, 
Unguris1997} The surprisingly good 
results of the asymptotic approximation are also confirmed by first-principles 
calculations.\cite{ Lang1993, Nordstrom1994, Kudrnovsky1994, Lang1996}

The exact expression of the interlayer exchange coupling may be expanded in a 
series of powers of $1/D$. The first non-zero term is the $D^{-2}$ term; it
corresponds to the asymptotic approximation. The higher order terms (varying 
like $D^{-n}$ with $n \ge 3$) are called the preasymptotic corrections.
 
The overall success of the asymptotic approximation
in explaining the experimental results obtained 
at fairly low spacer thicknesses 
is {\em a priori\/} surprising and raises the question of a quantitative 
determination of the range of validity of the asymptotic approximation. In 
addition, there are also some cases where the asymptotic approximation fails 
in providing an accurate description of the interlayer coupling for small 
spacer thicknesses. 

The inadequacy of the asymptotic approximation for an accurate description 
of the interlayer exchange coupling is best illustrated by the Co/Cu/Co(001) 
system. The ``exact'' calculations (i.e., not relying on the asymptotic 
approximation) show that the $D^{-2}$ decay law of short period oscillation 
is not obeyed until the spacer thickness reaches 
approximately 20~AL.\cite{ Drchal1996}
In addition, the amplitude of the long period oscillation 
for Co/Cu/Co(001) with thick Co layers as calculated from the asymptotic 
approximation\cite{ Bruno1995b, Mathon1995, Lee1995, Stiles1996, 
Albuquerque1996, 
Mathon1997} is typically at least one order of magnitude too small as compared 
to the one obtained from calculations which do not rely on the asymptotic 
approximation.\cite{ Lang1996, Drchal1996} The inadequacy of the asymptotic 
approximation to describe accurately the long period oscillation of the 
Co/Cu/Co(001) system with thick Co layers is also illustrated by the fact that 
the corresponding amplitude, as well as the period itself depends on the Cu 
thickness range which is used to determine it\cite{ Drchal1996, 
Kudrnovsky1998} 
and usually 
does not agree with the one calculated from the bulk Cu Fermi 
surface.\cite{ Bruno1998}

The present paper is devoted to a detailled discussion of the validity of the 
asymptotic approximation and of the preasymptotic corrections. It is organized 
as follows. First, in Sec.~\ref{ sec:asympt} I recall the assumptions 
underlying the asymptotic approximation, and the results that it yields. 
Next, in Sec.~\ref{ sec:preasympt} I present a systematic method for 
calculating the preasymptotic corrections to arbitrary order and their 
temperature dependence, and we carry out 
explicitely the calculation of the first preasymptotic correction, i.e., 
the $D^{-3}$ term. The Sec.~\ref{ sec:discuss} is devoted to the discussion 
of the results with emphasis on the Co/Cu/Co(001) system, and concluding 
remarks are given in Sec.~\ref{ sec:concl}.

\section{Asymptotic approximation}\label{ sec:asympt}

The general form of the interlayer exchange coupling is\cite{ Bruno1995}
\begin{eqnarray}\label{ eq:exact}
J(D,T) &=& -\mbox{Im} \int_{-\infty}^{+\infty} \!\!{\rm d}\varepsilon \, 
f(\varepsilon ,T) \nonumber \\
&& \times  \int {\rm d}^2{\bf k}_\| \, M(\varepsilon , {\bf k}_\|) \,
{\rm e}^{{\rm i}q(\varepsilon , {\bf k}_\|)D}.
\end{eqnarray}
In this expression, $q(\varepsilon , {\bf k}_\| ) \equiv k_z^+ - k_z^-$ is the 
difference between wave-vectors of an electron propagating through the spacer  
layer in the $+z$ and $-z$ directions (the $z$ axis is taken perpendicular to 
the layer plane). Here a single contribution has been considered; in the 
general 
case there would be several such contributions, due to multiple bands in the 
spacer materials and to higher harmonics (i.e., higher order terms in an 
expansion in powers of ${\rm e}^{{\rm i} q(\varepsilon ,{\bf k}_\| )}$) but 
the calculation of the various contributions is exactly the same; 
thus, for the
sake of simplicity, a single contribution is considered here. As explicitely 
indicated, $q(\varepsilon , {\bf k}_\| )$ varies with the energy $\varepsilon$
and with the in-plane wave-vector ${\bf }_\|$. The other factor 
in the integrand 
of Eq.~(\ref{ eq:exact}) are the Fermi-Dirac function $f(\varepsilon ,T)$ and 
the complex amplitude $M(\varepsilon , {\bf k}_\| )$ which depends on the 
spin-asymmetry of the reflection coefficients at the spacer-ferromagnet 
interfaces.\cite{ Bruno1995} The range of the integration over ${\bf k}_\|$ in 
Eq.~(\ref{ eq:exact}) is the surface Brillouin zone corresponding to the 
crystalline orientation of the layers.

The asymptotic approximation is based upon the observation that, because of 
the strong variation of the Fermi-Dirac function at the Fermi energy, and 
because of the rapid variation with $\varepsilon$ and ${\bf k}_\|$ of the 
exponential factor, the behavior of Eq.~(\ref{ eq:exact}) at large $D$ is 
dominated by the contribution of states on the Fermi surface, such that the 
spanning vector of the Fermi surface, $q(\varepsilon ,{\bf k}_\| )$, is 
stationary with respect to ${\bf k}_\|$.\cite{ Bruno1991} In the general case, 
there may several such stationary spanning vectors, each of them giving rise
to an oscillatory component of the interlayer exchange coupling, but for 
simplicity, I shall consider the case of a single component; 
the generalization 
to the case of multiple components is immediate.

The in-plane wave-vector ${\bf k}_\|^\star$ corresponding to the stationary 
spanning vector of the Fermi surface $q^\star$ is taken as the origin for 
${\bf k}_\|$ and the Fermi level is taken as the origin for $\varepsilon$.

The variation of $M(\varepsilon ,{\bf k}_\| )$ with $\varepsilon$
and ${\bf k}_\|$ is neglected, i.e., we assume
\begin{equation}\label{ eq:approx1}
M(\varepsilon ,{\bf k}_\| ) \approx M_{0;0,0}
\end{equation}
where $M_{0;0,0}$ is a constant (the motivation for the choice of the 
notations 
will appear clearly below). Further, we expand 
$q(\varepsilon ,{\bf k}_\| )$ near $q^\star$ as
\begin{equation}\label{ eq:approx2}
q(\varepsilon , {\bf k}_\|) \approx q^\star + \frac{2\varepsilon}{\hbar v_F}
- \frac{{k_x}^2}{\kappa_x} - \frac{{k_y}^2}{\kappa_y} ,
\end{equation}
where $v_F$ is the Fermi velocity, and $\kappa_x$ and $\kappa_y$ 
the curvature 
radii of the Fermi surface corresponding to ${\bf k}^\star_\|$ 
(the $x$ and $y$ 
axes are chosen so as to eliminate the term proportional to $k_x k_y$). 
Finally, since only 
the neighbohood of ${\bf k}_\|^\star$ contributes significantly to 
the integral,
we extend the integration range for $k_x$ and $k_y$ from $-\infty$ to 
$+\infty$. Then, we obtain easily\cite{ Bruno1995}
\begin{equation}
J(D,T) \approx \mbox{Im} \left[ \frac{{\rm e}^{{\rm i}q^\star D}}{D^2}
\, A_0(DT) \right]
\end{equation}
with the complex amplitude determined by
\begin{eqnarray}
A_0(DT) &\equiv & \frac{\pi}{2} \hbar v_F\, (\kappa_x)^{1/2}\, 
(\kappa_y)^{1/2}\, M_{0;0,0} \nonumber \\
&& \times F_0\left( \frac{2\pi\, k_BT\, D}{\hbar v_F} \right) .
\end{eqnarray}
The temperature dependence is given by the function
\begin{equation}
F_0(x) \equiv \frac{x}{\sinh x} .
\end{equation}

The predictions of the asymptotic approximation are summarized as follows: 
(i) at $T=0$ the coupling is given by a periodic function of $D$ multiplied by 
a decay factor $D^{-2}$, (ii) the phase of the oscillations is 
independent of the temperature, (iii) the complex amplitude factor  
varies with spacer thickness and temperature only as a function of the product 
$DT$.

\section{Calculation of the preasymptotic corrections}\label{ sec:preasympt}

Let us compute the preasymptotic corrections that appear when we release 
the approximations made in the previous section. First, instead of assuming 
that $M(\varepsilon ,{\bf k}_\| )$ is a constant, we expand it around 
$\varepsilon_F \equiv 0$ and ${\bf k}_\|^\star \equiv 0$ as
\begin{equation}
M(\varepsilon , {\bf k}_\|) \equiv  \sum_{p,q,r \ge 0} 
 M_{p;q,r}\ 
\varepsilon ^p \,
{k_x}^q \, {k_y}^r ,
\end{equation}
which defines the expansion coefficients $M_{p;q,r}$. In addition, we extend 
the expansion of $q(\varepsilon , {\bf k}_\|)$ beyond the first order in 
$\varepsilon$ and the second order in $k_x$ and $k_y$:
\begin{equation}
q(\varepsilon , {\bf k}_\|) = q^\star + \frac{2\varepsilon}{\hbar v_F}
- \frac{{k_x}^2}{\kappa_x} - \frac{{k_y}^2}{\kappa_y} + 
Q(\varepsilon , {\bf k}_\|)
\end{equation}
\begin{equation}
Q(\varepsilon , {\bf k}_\|) \equiv \sum_{2s+t+u > 2} Q_{s;t,u}\ 
\varepsilon^s\,
{k_x}^t \,{k_y}^u
\end{equation}
which defines the expansion coefficients $Q_{s;t,u}$. The factor 
${\rm e}^{{\rm i}qD}$ (the $\varepsilon$ and ${\bf k}_\|$ arguments will be 
dropped in the following) in Eq.~(\ref{ eq:exact}) is then rewritten as
\begin{eqnarray}
{\rm e}^{{\rm i}qD} &=& {\rm e}^{\rm iq^\star D} \,
\exp\left( \frac{{\rm i} \varepsilon D}{\hbar v_F}\right) \,
\exp \left( \frac{-{\rm i}k_xD}{\kappa_x} \right) \,
\exp \left( \frac{-{\rm i}k_yD}{\kappa_y} \right) \nonumber \\
&& \times \, \sum_{n\ge 0} \frac{(iQD)^n}{n!} .
\end{eqnarray}
Then the product $M(\varepsilon , {\bf k}_\|$ with the last factor of the 
above equation is expanded in powers of $\varepsilon$, $k_x$ and $k_y$ as 
follows
\begin{equation}
M \sum_{n\ge 0} \frac{(iQD)^n}{n!} = 
\sum_{p,q,r \ge 0} C_{p;q,r}(D) \ \varepsilon^p \, {k_x}^q \, {k_y}^r ,
\end{equation}
which defines the new expansion coefficients $C_{p:q,r}(D)$. Their explicit 
expression in terms of the coefficients $M_{p;q,r}$ and $Q_{s;t,u}$ can be 
obtained by a straighforward term-by-term identification.

Inserting the above expressions in Eq.~(\ref{ eq:exact}), we obtain
\begin{eqnarray}
J(D,T) &=& -\mbox{Im} \left[ {\rm e}^{{\rm i}q^\star D} \frac{ }{ } \right.
\sum_{p,q,r \ge 0} C_{p;q,r}(D)  \nonumber \\
&& \times 
\int_{-\infty}^{+\infty} \!\! {\rm d}\varepsilon \, \varepsilon^p \,
f(\varepsilon ,T) \, \exp\left( \frac{{\rm i}\varepsilon D}{\hbar v_F}\right)
\nonumber \\
&&\times \int_{-\infty}^{+\infty} \!\! {\rm d}k_x \, {k_x}^q \, 
\exp\left(\frac{-{\rm i}{k_x}^2D}{\kappa_x} \right)
\nonumber \\
&&\times \left. \int_{-\infty}^{+\infty} \!\! {\rm d}k_y \, {k_y}^r \, 
\exp\left(\frac{-{\rm i}{k_y}^2D}{\kappa_y} \right) \right]
\end{eqnarray}
The above expression is non-zero only if $q$ and $r$ are even. Performing the
integrations as explained in the Appendix, we obtain 
\begin{eqnarray}\label{ eq:preas1}
J(D,T) &=& \mbox{Im} \left[ \frac{{\rm e}^{{\rm i}q^\star D}}{D^2}\,
\frac{\pi}{2} \, \hbar v_F \, (\kappa_x)^{1/2} \, (\kappa_y)^{1/2} \right.
\nonumber \\
&& \times \sum_{p,q,r \ge 0} C_{p;2q,2r}(D)\ \frac{1}{D^{p+q+r}} \nonumber \\
&& \times \ \frac{(-1)^{q+r}\, {\rm i}^{p+q+r}}{2^{p+q+r}} 
\ p!\,\frac{(2q+1)!!}{2q+1}\, \frac{(2r+1)!!}{2r+1} 
\nonumber \\
&& \times \ \left. (\hbar v_F)^p\, {\kappa_x}^q \, {\kappa_y}^r \,
F_p\left( \frac{2\pi\, k_BT\, D}{\hbar v_F} \right) \right] ,
\end{eqnarray}
with 
\begin{equation}
(2n+1)!! \equiv \frac{(2n+1)!}{2^n\, n!} ,
\end{equation}
and where the functions $F_n(x)$ governing the temperature dependence of the 
coupling are defined by
\begin{equation}\label{ eq:F_n}
F_n(x) \equiv \frac{(-1)^n}{n!}\, x^{n+1}\, 
\frac{{\rm d}^n}{{\rm d}x^n} \left( \frac{1}{\sinh x}\right) .
\end{equation}
Finally, we reorder the terms of Eq.~(\ref{ eq:preas1}) so as to 
express it as 
a series in powers of $1/D$:
\begin{equation}\label{ eq:preas2}
J(D,T) = \mbox{Im} \left[ \frac{{\rm e}^{{\rm i}q^\star D}}{D^2} 
\sum_{n\ge 0} \frac{A_n(DT)}{D^n} \right] .
\end{equation}
The expressions of the functions $A_n(DT)$ in terms of the coefficients 
$M_{p;q,r}$ and $Q_{s;t,u}$ are obtained from 
Eq.~(\ref{ eq:preas1}) and from the explicit expression of the expansion 
coefficients $C_{p;q,r}(D)$ after tedious but straightforward algebraic 
manipulations. Below, I give the explicit expression of the asymptotic 
approximation term $A_0$ and of the first preasymptotic correction term $A_1$:
\begin{mathletters}
\begin{eqnarray}
A_0 &=& \frac{\pi}{2}\, \hbar v_F\, (\kappa_x)^{1/2}\, (\kappa_y)^{1/2}\,
M_{0;0,0}\, F_0 \\
{ }^{ } \nonumber \\
A_1 &=& \frac{\pi}{2} \, \hbar v_F\, (\kappa_x)^{1/2}\, (\kappa_y)^{1/2}
\nonumber \\
& \times & \left\{ -\frac{{\rm i}}{2} (M_{0;2,0}\, \kappa_x +
M_{0;0,2} \, \kappa_y)\, F_0 \right .\nonumber \\
&& - \frac{3}{4} [ ( M_{0;0,0}\, Q_{0;4,0} + M_{0;1,0}\, Q_{0;3,0})\, 
{\kappa_x}^2
\nonumber \\
&& \ \ \, + ( M_{0;0,0}\, Q_{0;0,4} + M_{0;0,1}\, Q_{0;0,3})\, 
{\kappa_y}^2 ]\, F_0 
\nonumber \\
&& - \frac{1}{4} (M_{0;0,0}\, Q_{0;2,2} + M_{0;1,0}\, Q_{0;1,2} 
\nonumber \\
&&\ \ \,+M_{0;0,1}\, Q_{0;2,1})\, \kappa_x \, \kappa_y\, F_0 \
+ \frac{i}{2} M_{1;0,0} \,\hbar v_F \, F_1
\nonumber \\
&& + \frac{1}{4} [( M_{0;0,0}\, Q_{1;2,0} + M_{0;1,0}\, Q_{1;1,0}) \,
\hbar v_F \, \kappa_x 
\nonumber \\
&& \ \ \, + (M_{0;0,0}\, Q_{1:0,2} + M_{0;0,1}\, Q_{1;0,1})\, \hbar v_F\, 
\kappa_y ] \, F_1
\nonumber \\
&& \left. \! - \frac{1}{2} M_{0;0,0}\, Q_{2;,0,0}\, 
(\hbar v_F)^2 \,F_2 \right\} . \label{ eq:A_1}
\end{eqnarray}
\end{mathletters}

\section{Discussion}\label{ sec:discuss}

\subsection{Symmetry considerations}\label{ sec:sym}

As discussed in detail in Ref.~\onlinecite{ Bruno1995}, the various critical 
points ${\bf k}_\|^\star$ of the two-dimensional Brillouin zone giving the 
oscillatory contributions to the interlayer 
exchange coupling can classified according to their symmetry: following the 
terminology introduced in Ref.~\onlinecite{ Bruno1995}, 
critical points corresponding to high-symmetry points of the two-dimensional 
Brillouin zone are termed {\em essential\/}, critical points located on 
high-symmetry lines are termed {\em semi-essential\/}, while critical points 
possessing no particular symmetry are termed {\em accidental\/}. 

The symmetry considerations impose a number of restrictions on the c
oefficients 
$M_{p;q,r}$, $Q_{s;t,u}$ and $C_{p;q,r}$. These are given below:

(i) For an {\em essential\/} critical 
point, we have $M_{p;q,r}=0$, $Q_{p;q,r} = 0$ and $C_{p;q,r} = 0$ if 
$q$ or $r$ is odd. If in addition, the critical point corresponds to 
a rotation 
axis of order equal to or larger than 3, then $\kappa_x = \kappa_y$, 
$M_{p;2q,2r} = M_{p;2r,2q}$, $Q_{p;2q,2r} = Q_{p;2r,2q}$, and $C_{p;2q,2r} = 
C_{p;2r,2q}$.

(ii) For a {\em semi-essential\/} critical point (without restriction, 
we chose 
the $y$ axis parallel to the high-symmetry line), we have $M_{p;q,r}=0$, 
$Q_{p;q,r} = 0$ and $C_{p;q,r} = 0$ if $q=0$.

(iii) For an {\em accidental\/} critical point, no restriction applies.

\subsection{Preasymptotic corrections at $T=0$}\label{ sec:T=0}

If we retain only the first preasymptotic correction term in 
Eq.~(\ref{ eq:preas2}), the coupling may be expressed as
\begin{equation}
J \approx \mbox{Im} \left[ \frac{{\rm e}^{{\rm i}q^\star D}}{D^2} \, 
A_0 \, \left( 1+ \frac{\Delta}{D} \right) \right] ,
\end{equation}
where the complex length
\begin{equation}
\Delta \equiv \frac{A_1}{A_0} 
\end{equation}
characterizes the preasymptotic correction. Thus, for $D \gg |\Delta |$, 
one has 
\begin{equation}
J \approx \frac{|A_0|}{D^2} \, 
\left( 1 + \frac{|\Delta |\cos \delta }{D}\right)
\, \sin \left( q^\star D + \phi + \frac{|\Delta | \sin\delta }{D} \right) ,
\end{equation}
where $\phi$ and $\delta$ are respectively the arguments of 
$A_0$ and $\Delta$.
From the above equation, it appears that the real part of $\Delta$ yields 
a correction of the amplitude of the oscillatory coupling; depending on 
whether 
$\cos\delta$ is positive or negative, one has an increase or a decrease of the
amplitude. On the other hand, 
the imaginary part of $\Delta$ contributes to a correction of the phase of the 
oscillation; i.e., for low spacer thickness, the apparent period of the 
oscillations differs from the asymptotic one\cite{ note}
\begin{equation}
\Lambda^\star \equiv \frac{2\pi}{| q^\star |} 
\end{equation}
and becomes 
\begin{equation}
\Lambda_{\mbox{\scriptsize app}} \approx \Lambda^\star \, 
\left(1- \frac{|\Delta|\sin\delta}{q^\star D^2} \right) ;
\end{equation}
depending on whether $\sin\delta$ and $q^\star$ are of the same sign or not, 
one as a decrease or an increase of the oscillation period.

Finally, the above discussion shows that a quantitative criterion for the 
validity of the asymptotic approximation is given by two following conditions
\begin{mathletters}
\begin{eqnarray}
D &\gg& |\Delta | \cos\delta , \\
D &\gg& \left( \frac{|\Delta |\, |\sin\delta|}{|q^\star |}\right)^{1/2} .
\end{eqnarray}
\end{mathletters}

\begin{figure}[h]
\begin{center}
\epsfxsize=8cm
\epsffile{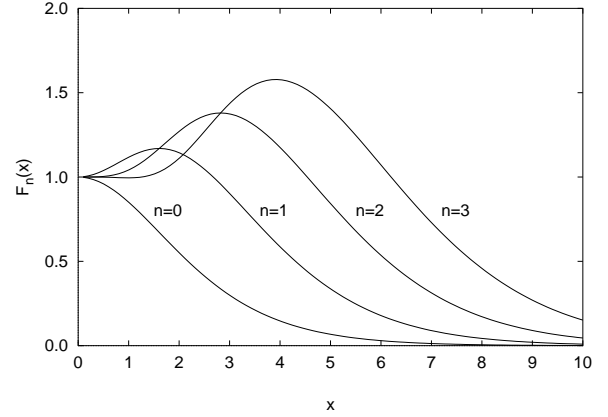}
\end{center}
\vspace*{-\baselineskip}
\caption{Plot of the functions $F_n(x)$ with $n = 0$ -- 3.}
\label{ fig:F_n}
\end{figure}

\subsection{Temperature dependence of the preasymptotic 
corrections}\label{ sec:T>0}

The temperature dependence of the coupling is governed by the functions 
$F_n(x)$ defined by Eq.~(\ref{ eq:F_n}). The functions $F_n(x)$ with 
$n=0$ -- 3 are given explicitely below:
\begin{mathletters}
\begin{eqnarray}
F_0 (x) &=& \frac{x}{\sinh x} \\
F_1(x) &=& \frac{x^2}{\sinh^2 x}\cosh x \\
F_2(x) &=& \frac{x^3}{\sinh^3 x} \left( 1+ \frac{\sinh^2 x}{2} \right) \\
F_3(x) &=& \frac{x^4}{\sinh^4 x} \cosh x 
\left( 1 + \frac{\sinh^2 x}{6} \right) .
\end{eqnarray}
\end{mathletters}
They are displayed on Fig.~\ref{ fig:F_n}. The functions $F_n(x)$ have the 
following general properties: 
\begin{mathletters}
\begin{eqnarray}
F_n(x) &\approx & 1 + \beta_n x^2  \mbox{ for } x \ll 1 , \\
F_n(x) &\approx & \frac{2}{n!}\, x^{n+1} \, {\rm e}^{-x} 
\mbox{ for } x \gg n+1 ,
\end{eqnarray}
\end{mathletters}
where the constants $\beta_n$ are given by
\begin{mathletters}
\begin{eqnarray}
\beta_0 &=& - \frac{1}{6} , \\
\beta_1 &=& \frac{1}{6} , \\
\beta_n &=& 0 \ \mbox{ for } n \ge 2 ;
\end{eqnarray}
\end{mathletters}
$F_n(x)$ with $n \ge 1$ has a maximum at
\begin{mathletters} 
\begin{eqnarray}
x_n^\star &\approx& n+1 , \ \mbox{ and } \\
F_n(x_n^\star ) &\approx& \frac{2}{n!} \left(\frac{n+1}{{\rm e}}\right)^{n+1}
\end{eqnarray}
\end{mathletters}

Because of the marked difference between $F_0(x)$ and the functions $F_n(x)$ 
with $n\ge 1$, the temperature dependence of the preasymptotic corrections 
is in general not the same as the one of the asymptotic term. In other 
words, the parameter $\Delta$ characterized the preasymptotic corrections 
defined in Sec.~\ref{ sec:T=0} generally varies with the temperature; 
thus, the extension of the asymptotic regime may depend on the temperature. 
As expected and appears clearly from Eq.~(\ref{ eq:A_1}), the temperature 
dependence of $\Delta$ 
arises from the terms proportional to $F_1$ and $F_2$, 
i.e. from the corrections 
due to the $\varepsilon$ variation of $M(\varepsilon , {\bf k}_\| )$ and 
$Q(\varepsilon , {\bf k}_\| )$.

\subsection{Comparison with previous work}

The question of preasymptotic corrections has been addressed previously 
by Mathon {\em et al.\/}\cite{ Mathon1995, Albuquerque1996, Mathon1997} 
They have considered the correction due to the energy dependence of 
the argument of $M(\varepsilon ,{\bf k}_\| )$. This correspond to assume 
that the only non-zero correction parameter is $M_{1;0,0}$ and that the 
ratio $M_{1;0,0}/M_{0;0,0}$ is purely imaginary. In this case, they find 
that the preasymptotic corrections modify the amplitude and the temperature 
dependence of the coupling. This conclusion is in agreement with the result 
of the present paper, if we restrict our analysis to the same assumptions. 
However, even within the restrictive assumptions, the method used by Mathon 
{\em et al.\/} is correct only to the order 
$D^{-3}$ because the the terms order $D^{-n}$ with ($n \ge 4$) which are 
neglected are of the same magnitude as the ones which are included.

Furthermore, as the present analysis shows, there are other 
sources of preasymptotic corrections, for example the ${\bf k}_\|$ variation 
of $M(\varepsilon ,{\bf k}_\| )$, that {\em a priori\/}, are equally 
important as 
the ones considered by Mathon {\em et al.\/}

\subsection{Discussion of the Co/Cu/Co(001) case}

We shall now focus on the particular case of the Co/Cu/Co(001) system, 
which is often considered as a model system for the problem of interlayer 
exchange coupling. 

\subsubsection{Summary of the available data}

Let us first recall the predictions of the asymptotic approximation for this 
system. The most striking result of the asymptotic approximation 
here is 
that the interlayer exchange coupling comprises a long-period oscillation 
originating for the center $\overline{\Gamma}$ of the two-dimensional 
Brillouin zone ({\em essential\/} critical point) and of a short-period 
oscillation originating 
from 4 equivalent {\em semi-essential\/} critical points located on 
the $\overline{\Gamma}$--$\overline{X}$ lines;\cite{ Bruno1991, Bruno1995}
this prediction is 
confirmed unambiguously both by experiments\cite{ Johnson1992} and by 
``exact'' (i.e., not relying on the asymptotic approximation) 
calculations,\cite{ Lang1993, Nordstrom1994, Kudrnovsky1994, Lang1996, 
Drchal1996} and the 
periods are in good agreement with the ones predicted by the asymptotic 
approximation. In the following, quantities related to the short-period and 
long-period components will labelled by $S$ and $L$ indices, respectively.

Quite generally, the function $M(\varepsilon ,{\bf k}_\| )$ determining the 
strength of the coupling is given by\cite{ Bruno1993, Stiles1993, Bruno1995} 
\begin{equation}
M(\varepsilon ,{\bf k}_\| ) \equiv \frac{\left(  r^\uparrow - 
r^\downarrow \right)^2}{8\pi^3} ,
\end{equation}
where $r^\uparrow$ ($r^\downarrow$) is the reflection coefficient
for electrons with the spin parallel (antiparallel) to the majority spin of 
Co. Since the majority spin band structure of Co is very close to the Cu 
band structure, one has $r^\uparrow \approx 0$ and it is sufficient to 
consider $r^\downarrow$.

For the short period oscillation, one has total reflection for minority 
spin electron, due to a local gap (for the bands of relevant symmetry) 
in the minority spin band structure of 
Co, i.e., $|r^\downarrow_S| =1$.\cite{ Mathon1995, Lee1995, Stiles1996, 
Albuquerque1996, Mathon1997} As a consequence, the amplitude of the 
short period oscillation is large and depends only very weakly upon the 
C thickness. These facts are confirmed by ``exact'' calculations.\cite{ 
Lang1993, Nordstrom1994, Kudrnovsky1994, Lang1996, Drchal1996} The period 
of oscillation obtained from an ``exact'' calculation, 
($\Lambda_S = 2.53$~AL)\cite{ Kudrnovsky1998} is in excellent agreement with 
the one calculated within the asymptotic approximation with the same set of 
potential parameters ($\Lambda_S = 2.50$~AL);\cite{ Bruno1998} furthermore, 
the apparent period agrees very well with the asymptotic one down to Cu 
thicknesses as low a 5~AL. On the other 
hand, as appears very clearly from the Fig.~3a of 
Ref.~\onlinecite{ Drchal1996}, 
the asymptotic $D^{-2}$ decay law is satisfied only for Cu spacer thicknesses 
larger 20~AL; in practice, this means that for a quantitative comparison 
with experiment 
results (which are usually obtained for spacer ticknesses smaller than 20~AL), 
the amplitude calculated from the asymptotic approximation is inappropriate.

For the long period oscillation, the situation is quite different. Here, the 
reflection coefficient $r^\downarrow_L$ is quite small for a semi-infinite 
Co layer ($|r^\downarrow_L | 
\approx 0.15$)\cite{ Bruno1995, Bruno1995b, Lee1995, Stiles1996, 
Dederichs1997} 
which leads to a very small amplitude for the long-period oscillatory 
component; all theoretical calculation, whether ``asymptotic''
\cite{ Bruno1995b, Mathon1995, Lee1995, Stiles1996, Albuquerque1996, 
Mathon1997} or ``exact''\cite{ Lang1993, Nordstrom1994, Kudrnovsky1994, 
Lang1996, Drchal1996} agree on this point. In addition, there is a strong 
variation of the amplitude of the long period oscillation 
with Co layer thickness, which is due to quantum interferences in 
the Co layer,\cite{ Bruno1993b} a prediction which has been confirmed 
experimentally\cite{ Bloemen1994} and by ``exact'' theoretical 
calculations.\cite{ Nordstrom1994, Kudrnovsky1994, Lang1996, Drchal1996, 
Dederichs1997} 
However, although both asymptotic\cite{ Bruno1995b, Mathon1995, Lee1995, 
Stiles1996, Albuquerque1996, Mathon1997} and ``exact''\cite{ Nordstrom1994, 
Kudrnovsky1994, Lang1996, Drchal1996} calculations agree on the fact 
that the long period oscillation has a very weak amplitude for thick 
Co layers, 
they disagree on the value of the amplitude: while ``exact'' calculation yield 
an amplitude of the long period oscillation of the order of 10~\%\ of the 
short period amplitude,\cite{ Nordstrom1994, Lang1996, Drchal1996} asymptotic 
calculations yield a ratio of the order of 1~\%\ only.\cite{ Mathon1995, 
Lee1995, Stiles1996, Albuquerque1996, Mathon1997} In addition, 
the value of the 
long period coupling obtained from ``exact'' calculations, 
$\Lambda_L = 5.09$~AL,\cite{ Kudrnovsky1998} differs markedly from the one 
calculated from the asymptotic approximation with the same set of potential 
parameters, $\Lambda_L = 6.49$~AL,\cite{ Bruno1998} and depends on the 
thickness 
range which is used to determine it as is seen from Fig.~2 of 
Ref.~\onlinecite{ Drchal1996}.

So, our purpose is to explain why the asymptotic approximation works well for 
some aspects of the coupling and departs markedly from ``exact'' calculations 
for some other aspects.

For both the long-period and short-period contribution, the preasymptotic 
corrections due to the $\varepsilon$ and ${\bf k}_\|$ dependence of 
$Q(\varepsilon ,{\bf k}_\| )$ is unimportant, so that we shall focus on the 
corrections due the variation of $M(\varepsilon ,{\bf k}_\| )$. Thus, 
by taking the symmetry considerations mentioned in Sec.~\ref{ sec:sym}, the 
corrections are given by
\begin{mathletters}
\begin{eqnarray}
\Delta_L &=& -{\rm i}\, \frac{M_{0;2,0}^L}{M_{0;0,0}^L} \, \kappa_x^L + 
\frac{{\rm i}}{2} \, \frac{M_{1;0,0}^L}{M_{0;0,0}^L} \, \hbar v_F^L \, 
\frac{F_1^L}{F_0^L} , \\
\Delta_S &=& -\frac{{\rm i}}{2} 
\left( \frac{M_{0;2,0}^S}{M_{0;0,0}^S} \, \kappa_x^S + 
\frac{M_{0;0,2}^S}{M_{0;0,0}^S} \, \kappa_y^S \right) \nonumber \\
&&+ \frac{{\rm i}}{2} \, \frac{M_{1;0,0}^S}{M_{0;0,0}^S} \, \hbar v_F^S \, 
\frac{F_1^S}{F_0^S} ,
\end{eqnarray}
\end{mathletters}
respectively.
Let us now recall that within elementary models such as the free-electron 
model or the single-band tight-binding model, the reflection coefficient 
for a semi-infine barrier is 
either a real number of module smaller than 1 (partial reflection) or a 
complex number of module equal to 1 (total reflection). As shown in 
Ref.~\onlinecite{ Bruno1995}, this remains approximately true for realistic 
multi-band systems under rather general circumstances. 

\subsubsection{Short-period oscillation}

Thus, for the short period oscillation, 
only the phase of the reflection coefficient varies with $\varepsilon$ and 
${\bf k}_\|$, and as a consequence, $\Delta_S$ is a real number. Hence, 
according to the discussion of Secs.~\ref{ sec:T=0} and \ref{ sec:T>0}, the 
preasymptotic correction affects only the amplitude of the coupling and its
temperature dependence, while the period is not affected by the preasymptotic 
corrections; these conclusions are in agreement with the results summarized 
above. Similar conclusions have been obtained by Mathon 
{\em et al.\/}\cite{ Mathon1997} who have considered the preasymptotic 
correction associated with the energy dependence $M(\varepsilon , 
{\bf k}_\| )$, 
i.e., with $M_{0;0,1}^S$
and neglected the one associated with its ${\bf k}_\|$ dependence, i.e., with 
$M_{0;2,0}^S$ and $M_{0;0,2}^S$;
indeed in view of the similarity of their corrected results (see Fig.~13 of 
Ref.~\onlinecite{ Mathon1997}) with those of ``exact'' calculations (see 
Fig.~3 of Ref.~\onlinecite{ Drchal1996}), it seems that the $M_{0;0,1}^S$ term 
already accounts for a large part of the total preasymptotic correction. 
Further 
work would be needed to assess the importance of the $M_{0;2,0}^S$ and 
$M_{0;0,2}^S$ terms.

\subsubsection{Long-period oscillation}

For the long-period oscillation, on the other hand, since $r^\downarrow_L$ is 
real, only the magnitude varies with $\varepsilon$ and ${\bf k}_\|$ and the 
phase is constant; thus, $\Delta_L$ is purely imaginary. As discussed above, 
this leads to an apparent shift of the period in the preasymptotic range, 
which is precisely what happens in this case. Let us attempt to estimate the 
value of $\Delta_L$. As seen from, e.g., Fig.~22 of 
Ref.~\onlinecite{ Bruno1995}, $r_\downarrow$ increases with 
decreasing energy. Furthermore, as seen from Fig.~2 of 
Ref.~\onlinecite{ Bruno1995b} and Fig.~2 of Ref.~\onlinecite{ Stiles1996}, 
$r^\downarrow$ increases very strongly with ${\bf k}_\|$ and full reflection 
is reached at a distance $0.1 \times \pi /a$ from $\overline{\Gamma}$; indeed, 
the low reflectivity arises only in a narrow window around $\overline{\Gamma}$. 
Taking these two contributions into account, we arrive at the result:
\begin{equation}\label{ eq:Delta_L}
\Delta_L \approx - {\rm i} \ \times \ 50 \mbox{ AL} .
\end{equation}
Because $q^\star_L$ is negative, this result implies that the preasymptotic 
correction will lead to an apparent oscillation period in the 
preasymptotic regime  that is shorter than the asymptotic one; this conclusion 
provides an consistent explanation for the discrepancy on the long 
oscillation 
period mentioned above. However, the preasymptotic regime here has an 
unusually 
large extension and asymptotic behavior is expected to hold only for 
$D \gg 50$~AL; thus, the analysis of Sec.~\ref{ sec:T=0}, which 
was limited to the lowest order in $1/D$, is certainly not sufficient 
to analyse 
the results obtained from experiments and from ``exact'' calculations 
and it is 
not surprising that not only the period itself, but also the amplitude of the 
long period oscillation predicted from the asymptotic approximation is 
inappropriate for spacer thicknesses $D \leq 50$~AL. 

Actually, 90\%\ of the 
result (\ref{ eq:Delta_L}) for $\Delta_L$ is due to the ${\bf k}_\|$ 
dependence 
of $r^\downarrow$, so that the unusually long preasymptotic regime for the 
long-period oscillation is to be attributed to the presence of a narrow 
window 
of low reflectivity near $\overline{\Gamma}$ in a region of otherwise total 
reflectivity, for minority spin. This pecularity can in turn be explained by
a rather simple following argument: The relevant band in the Cu is mostly of 
$p_z$ character. The reflectivity is due to the hybridization of the 
corresponding $p_z$ band in Co with the minority spin $d$ bands. At the center 
$\overline{\Gamma}$ of the two-dimensional Brillouin zone, only the 
$d_{3z^2-r^2}^\downarrow$ band is allow by symmetry to hybridize with the 
$p_z^\downarrow$ band, and 
hence to contribute to the reflectivity; however, because the 
$d_{3z^2-r^2}^\downarrow$ band is almost full, it yields only a weak 
reflection 
coefficient at the Fermi level, and total reflection is attained only $0.5$~eV 
below $\varepsilon_F$. But, as one moves away for $\overline{\Gamma}$, the 
$d_{xz}^\downarrow$ and $d_{yz}^\downarrow$ bands that lie close to the Fermi 
level are allowed to hybridize with the $p_z^\downarrow$ band, which yields 
a strong increase of $r^\downarrow$ and, eventually, total reflection.

\section{Conclusion}\label{ sec:concl}

I have presented a detailed discussion of the question of preasymptotic 
corrections for the interlayer exchange coupling. A systematic method for 
computing exactly the preasymptotic corrections 
to arbitrary order in $D^{-1}$ 
has been presented, and the explicit expression of the first correction 
term (of order $D^{-3}$) has been given. 

This method allows one to assess quantitatively the spacer thickness range in 
which the asymptotic approximation is expected to be reliable, and the one in 
which preasymptotic corrections should be taken into account. In the latter 
case, I have shown that the preasymptotic correction alters the amplitude 
and/or the apparent oscillation period, depending on the argument of the 
correction parameter $\Delta$. 

The case of Co/Cu/Co(001) has been discussed in detail. I have shown that 
most of the discrepancy between the asymptotic approximation and ``exact'' 
calculations for the system can be understood on the basis of the theory 
for preasymptotic corrections presented in this paper.

\appendix
\section*{}

\subsection*{Integration over energy}

We consider here the following integral
\begin{equation}
I_n(T) \equiv \int_{-\infty}^{+\infty} {\rm d}\varepsilon \, f(\varepsilon ,T) 
\, \varepsilon^n \, \exp \left( 2{\rm i} 
\varepsilon \alpha \right) .
\end{equation}
We can compute it by using the method of residues. The Fermi-Dirac function 
$f(\varepsilon ,T)$ has
poles for 
\begin{equation}
\varepsilon_p = {\rm i} \pi (2p+1)\, \beta^{-1}
\end{equation}
where $p$ is an integer and $\beta \equiv (k_BT)^{-1}$. 
The corresponding residues are equal to $-\beta^{-1}$. 
Closing the integration path in the upper half of the complex plane, we obtain
\begin{equation}
I_n(T) = -2 \left( \frac{{\rm i} \pi}{\beta} \right)^{\! n+1} \,
\sum_{p=0}^{+\infty} 
\left( 2p+1\right)^n \, {\rm e}^{-2\pi (2p+1) \alpha /\beta} .
\end{equation}
Finally, we obtain 
\begin{equation}
I_n(T) = I_n^0 \ F_n\left( \frac{2\pi \, \alpha}{\beta} \right)
\end{equation}
whith
\begin{equation}
I_n^0 = - {\rm i}^{n+1} \alpha^{-(n+1)} n!
\end{equation}
and
\begin{equation}
F_n(x) \equiv \frac{(-1)^n}{n!}\, x^{n+1}\, 
\frac{{\rm d}^n}{{\rm d}x^n} \left( \frac{1}{\sinh x}\right) .
\end{equation}

\subsection*{Integration over wave vector}

Here we calculate the following integral.
\begin{equation}
B_{2n} \equiv \int_{-\infty}^{\infty} {\rm d} k \, k^{2n} 
\exp \left(- {\rm i}\gamma k^2 \right) .
\end{equation} 
We can deform the integration path in the complex plane in such a way that the 
point $k=0$ be traversed along the direction of steepest 
descent.\cite{ Negele1988} The integral is then easily calculated, and we get:
\begin{equation}
B_{2n} = \left( \frac{-{\rm i}}{\gamma} \right)^{n+1/2} \, 
\frac{(2n+1)!!}{2^n \, (2n+1)} \, \sqrt{\pi} .
\end{equation}


\begin{references}
\bibitem[*]{e-mail} Electronic address: {\tt bruno@mpi-halle.de}

\bibitem{ Parkin1990} S.S.P.~Parkin, N.~More, and K.P.~Roche, 
Phys. Rev. Lett. {\bf 64}, 2304 (1990).

\bibitem{ Heinrich1994} K.B.~Hathaway, A.~Fert, P.~Bruno, D.T.~Pierce, 
J.~Unguris, R.J.~Celotta, and S.S.P.~Parkin, in {\em Ultrathin Magnetic 
Films\/}, vol.2, edited by B.~Heinrich and J.A.C.~Bland (Springer-Verlag, 
Berlin, 1994), chap. 2, p. 45.

\bibitem{ Slonczewski1994} J.C.~Slonczewski, J. Magn. Magn. Mater. {\bf 150},
13 (1994).

\bibitem{ Edwards1991} D.M.~Edwards, J.~Mathon, R.B.~Muniz, and M.S.~Phan, 
Phys. Rev. Lett. {\bf 67}, 493 (1991).

\bibitem{ Bruno1993} P.~Bruno, J. Magn. Magn. Mater. {\bf 121}, 248 (1993).

\bibitem{ Stiles1993} M.D.~Stiles, Phys. Rev. B {\bf 48}, 7238 (1993).

\bibitem{ Bruno1995} P.~Bruno, Phys. Rev. B {\bf 52}, 411 (1995).

\bibitem{ Bruno1991} P.~Bruno and C.~Chappert, Phys. Rev. Lett. {\bf 67}, 1602 
(1991); {\bf 67} 2592(E) (1991).

\bibitem{ Yafet1987} Y.~Yafet, Phys. Rev. B {\bf 36}, 3948 (1987).

\bibitem{ Bruno1992} P.~Bruno and C.~Chappert, Phys. Rev. B {\bf 46} 
261 (1992).

\bibitem{ Unguris1994} J.~Unguris, R.J.~Celotta, and D.T.~Pierce, 
J. Appl. Phys.
{\bf 75}, 6437 (1994).

\bibitem{ Unguris1997} J.~Unguris, R.J.~Celotta, and D.T.~Pierce, Phys. Rev. 
Lett. {\bf 79}, 2734 (1998).

\bibitem{ Lang1993} P.~Lang, L.~Nordstr\"om, R.~Zeller, and P.H. Dederichs, 
Phys. Rev. Lett. {\bf 71}, 1927 (1993).

\bibitem{ Nordstrom1994} L.~Nordstr\"om, P.~Lang, R.~Zeller, and 
P.H.~Dederichs,
Phys. Rev. B {\bf 50}, 13058 (1994).

\bibitem{ Kudrnovsky1994} J.~Kudrnovsk\'y, V.~Drchal, I.~Turek, and 
P.~Weinberger, Phys. Rev. B {\bf 50}, 16105 (1994).

\bibitem{ Lang1996} P.~Lang, L.~Nordstr\"om, K.~Wildberger, R.~Zeller, 
P.H.~Dederichs, and T.~Hoshino, Phys. Rev. B {\bf 53}, 9092 (1996).

\bibitem{ Drchal1996} V.~Drchal, J.~Kudrnovsk\'y, I.~Turek, and P.~Weinberger,
Phys. Rev. B {\bf 53}, 15036 (1996).

\bibitem{ Bruno1995b} P.~Bruno, J. Magn. Magn. Mater. {\bf 148}, 202 (1995).

\bibitem{ Mathon1995} J.~Mathon, M.~Villeret, R.B.~Muniz, J.~d'Albuquerque e 
Castro, and D.M. Edwards, Phys. Rev. Lett. {\bf 74}, 3696 (1995).

\bibitem{ Lee1995} B.~Lee and Y.C.~Chang, Phys. Rev. B {\bf 52}, 3499 (1995).

\bibitem{ Stiles1996} M.D.~Stiles, J. Appl. Phys. {\bf 79}, 5805 (1996).

\bibitem{ Albuquerque1996} J.~d'Albuquerque e Castro, J.~Mathon, M.~Villeret, 
and A.~Umerski, Phys. Rev. B {\bf 53}, R13306 (1996).

\bibitem{ Mathon1997} J.~Mathon, M.~Villeret, A.~Umerski, R.B.~Muniz, 
J.~d'Albuquerque e Castro, and D.M.~Edwards, Phys. Rev. B {\bf 56}, 11797 
(1997).

\bibitem{ Kudrnovsky1998} J.~Kudrnovsky (private communication).

\bibitem{ Bruno1998} P.~Bruno (unpublished).

\bibitem{ note} As usual,\cite{ Bruno1995, Bruno1991} 
since $q^\star$ is defined only within a multiple of 
$2\pi /d$ ($d$ being the interplane distance in the normal direction), we 
take aliasing into account by making the choice that yields $-\pi /d \le 
q^\star  \le +\pi /d$.

\bibitem{ Johnson1992} M.T.~Johnson, S.T.~Purcell, N.W.E.~McGee, R.~Coehoorn, 
J.~aan de Stegge, and W.~Hoving, Phys. Rev. Lett. {\bf 68}, 2688 (1992).

\bibitem{ Dederichs1997} P.H.~Dederichs, K.~Wildberger, and R.~Zeller, 
Physica B 
{\bf 237-238}, 239 (1997).

\bibitem{ Bruno1993b} P.~Bruno, Europhys. Lett. {\bf 23}, 615 (1993).

\bibitem{ Bloemen1994} P.J.H.~Bloemen, M.T.~Johnson, M.T.H.~van de Vorst, 
R.~Coehoorn, J.J.~de Vries, R.~Jungblut, J.~aan de Stegge, A.~Reinders, and 
W.J.M.~de Jonge, Phys. Rev. Lett. {\bf 72}, 764 (1994).

\bibitem{ Negele1988} J.W.~Negele and H.~Orland, {\em Quantum Many-Particle 
Systems\/} (Addison-Wesley, Redwood City, 1988), p.~121.

\end{references}
\end{document}